\documentclass[english,journal=jctcce,manuscript=article,email=true,etalmode=truncate,maxauthors=0,doi=true]{achemso}

\title{Slimmer Geminals For Accurate F12 Electronic Structure Models.}
\date{\today}

\usepackage{amsmath,amssymb}
\usepackage{booktabs}
\usepackage[version=4]{mhchem} % for \ce command for formatting chemical formulae
\usepackage{tablefootnote}

\graphicspath{{figures/}}
\usepackage{physics}
\usepackage{multirow}
\usepackage{xcolor}
\usepackage{caption}
\usepackage{subcaption}
\usepackage{xcolor}
\usepackage{bm}
\usepackage{eufrak} % for \mathfrak
\usepackage{yfonts} % for \textgoth
\usepackage[title,titletoc]{appendix}
\usepackage{diagbox}
\usepackage{multicol}
\usepackage{mhchem}

\usepackage[numbers,sort&compress,super]{natbib}
\usepackage{natmove}
\usepackage{hyperref} % hyperref must be placed before cleveref
\hypersetup{colorlinks=true, citecolor=blue, urlcolor=blue, linkcolor=blue}
\usepackage{cleveref}
  	\crefname{figure}{Figure}{Figures}
  	\crefname{table}{Table}{Tables}
  	\crefname{equation}{Eq.}{Eqs.}
  	\crefname{section}{Section}{Sections}
  	\crefname{subsection}{Section}{Sections}
  	\crefname{subsubsection}{Section}{Sections}
  	\crefname{algorithm}{Algorithm}{Algorithms}

\newcommand{\code}[1]{\texttt{#1}}
\newcommand\vartextvisiblespace[1][.5em]{%
  \makebox[#1]{%
    \kern.07em
    \vrule height.3ex
    \hrulefill
    \vrule height.3ex
    \kern.07em
  }% <-- don't forget this one!
}
\usepackage{todonotes}

\author{Samuel R. Powell}
\author{Kshitijkumar A. Surjuse}
\author{Bimal Gaudel}
\author{Edward F. Valeev}
\affiliation{Department of Chemistry, Virginia Tech, Blacksburg, Virginia, 24060}
\email{efv@vt.edu}

\begin{document}

\begin{abstract}
The Slater-type F12 geminal lengthscales originally tuned for the second-order M{\o}ller-Plesset F12 method are too large for higher-order F12 methods formulated using the SP (diagonal fixed-coefficient spin-adapted) F12 ansatz. The new geminal parameters reported herein reduce the basis set incompleteness errors (BSIE) of absolute coupled-cluster singles and doubles F12 correlation energies by a significant --- and increasing with the cardinal number of the basis --- margin. The effect of geminal reoptimization is especially pronounced for the cc-pV$X$Z-F12 basis sets (specifically designed for use with F12 methods) relative to their conventional aug-cc-pV$X$Z counterparts. The BSIEs of relative energies are less affected but substantial reductions can be obtained, especially for atomization energies and ionization potentials with the cc-pV$X$Z-F12 basis sets. The new geminal parameters are therefore recommended for all applications of high-order F12 methods, such as the coupled-cluster F12 methods and the transcorrelated F12 methods.
\end{abstract}

\maketitle

\section{Introduction}\label{sec:intro}
The notoriously slow basis set convergence of the electronic energies and other properties evaluated with correlated electronic structure models arises primarily because of the lack of many-electron cusps in the conventional Fock space basis of products of single-particle states. To reduce the basis set incompleteness error (BSIE), one can extrapolate the property of interest to the complete basis set limit using one of the families of systematically designed basis sets, namely the correlation-consistent bases\cite{VRG:dunning:1989:JCP,VRG:kendall:1992:JCP,VRG:peterson:2008:JCP} or the atomic natural orbital (ANO) bases\cite{VRG:almlof:1987:JCP,VRG:roos:2005:CPL,VRG:neese:2011:JCTC,VRG:zobel:2020:JCTC}. Although some basis set extrapolation formulas can rationalized\cite{VRG:klopper:2001:MP} there are many formulas in use that are largely ad hoc recipes\cite{VRG:feller:1992:JCP,VRG:helgaker:1997:JCP,VRG:halkier:1998:CPL,VRG:schwenke:2005:JCP}
whose reliability is only as good as the amount of available benchmark data. The bulk of the basis set extrapolation testing has concentrated on light elements and the correlation-consistent basis set family, though studies of the robustness of basis set extrapolation for the energies obtained with ANO bases and for heavy elements (where ANOs and basis sets containing effective core potentials (ECPs) are the dominant choices) are ongoing.\cite{VRG:granatier:2017:ACS,VRG:larsson:2022:ES,VRG:peterson:2003:JCP,VRG:peterson:2003:JCPa,VRG:peterson:2005:TCA,VRG:balabanov:2005:JCP,VRG:peterson:2007:JCP,VRG:figgen:2009:JCP,VRG:williams:2008:JCP,VRG:balabanov:2006:JCP,VRG:aoto:2017:JCTC,VRG:minenkov:2016:JCTC,VRG:cheng:2017:JCTC} 
% Correlation consistent basis sets exist for much of the periodic table (often with effective core potentials), but the authors are unaware of significant studies on basis set extrapolation with these basis sets. \cite{VRG:peterson:2003:JCP,VRG:peterson:2003:JCPa,VRG:peterson:2005:TCA,VRG:balabanov:2005:JCP,VRG:peterson:2007:JCP,VRG:figgen:2009:JCP}

The first-principles solution to the basis set problem is to use the so--called explicitly correlated models which contain Fock space basis functions that model the many-electron cusps by including explicit dependence on the interelectronic distances. Following the pioneering work of Hylleraas\cite{VRG:hylleraas:1929:ZP} and Slater\cite{VRG:slater:1928:PRa,VRG:slater:1928:PRb} many such models\cite{VRG:kong:2012:CR,VRG:hattig:2012:CR,VRG:ten-no:2012:TCA} have been developed targeting the high-precision simulation of small systems as well as applicability to larger systems.

There are several groups of explicitly-correlated approaches in use today:
\begin{itemize}
\item {\bf high-precision methods}: these involve unfactorizable many-particle integrals with most or all particles at once that are evaluated analytically; examples include Hylleraas-CI,\cite{VRG:ruiz:2021:AiQC} explicitly correlated Gaussian geminals\cite{VRG:mitroy:2013:RMP}, and free iterative complement interaction (ICI) approach of Nakatsuji\cite{VRG:nakatsuji:2004:PRL}.
\item {\bf configuration-space Jastrow factor}: 
these are methods involving unfactorizable many-particle integrals of most or all particles at once that are evaluated stochastically or avoided by serving as a trial wave function; examples include single- and multiconfiguration Slater-Jastrow wave functions in variational and diffusion quantum Monte-Carlo methods.\cite{VRG:foulkes:2001:RMP}
\item {\bf configuration-space transcorrelation}: this involves unfactorizable integrals with at most 3 particles that are evaluated analytically;\cite{VRG:handy:1969:JCP,VRG:boys:1969:PRSLA}
\item {\bf R12/F12 methods}: these involve unfactorizable integrals with up to 4 particles that are evaluated analytically (up to 2-particle integrals) or by the resolution-of-the-identity (3- and 4-particle integrals).\cite{VRG:klopper:2006:IRPC,VRG:kong:2012:CR,VRG:hattig:2012:CR,VRG:ten-no:2012:WCMS,VRG:ten-no:2012:TCA}
\end{itemize}

By only requiring the exact evaluation of (non-standard) 2-particle integrals, the F12 methods are most practical among all explicitly correlated methods, available routinely in several packages and with demonstrated applications to hundreds and thousands of atoms.\cite{VRG:pavosevic:2016:JCP,VRG:ma:2018:JCTC,VRG:kumar:2020:JCP,VRG:wang:2023:JCTC}

Other key advantages of modern F12 methods over most other explicitly correlated methods are that (a) no problem-specific nonlinear optimization of the parameters of the explicitly correlated terms is required and (b) the spin-dependence of the cusp conditions\cite{VRG:pack:1966:JCP} can be satisfied rigorously unlike the purely configuration-space methods\cite{VRG:huang:1998:JCP}. Following Ten-no\cite{VRG:ten-no:2004:CPL} the standard formulation of F12 methods uses a single Slater-type geminal (STG)
\begin{align}
\label{eq:stg}
f_\beta(r_{12}) \equiv & -\frac{\exp(-\beta r_{12}) }{ \beta}
\end{align}
to correlate every pair of electrons in the molecule. The recommended value of {\em inverse lengthscale} $\beta$ was pre-optimized for valence-only\cite{VRG:peterson:2008:JCP} and all-electron\cite{VRG:hill:2010:JCP} correlated computations with the most common orbital basis sets, but it is kept fixed in the course of the computation and no other adjustable parameters (linear or nonlinear) are associated with the explicitly correlated terms. The use of system-independent pre-optimized parameters should be contrasted with the other types of explicitly correlated methods, almost all of which involve optimization of adjustable nonlinear parameters of the explicitly correlated terms, which impacts their robustness. For example, to improve robustness Szenes et al. recently argued\cite{VRG:szenes:2024:FD} that simpler system-independent correlators are more likely to be useful in the context of transcorrelated methods, which is fully in the spirit of the F12 methods.

The lack of adjustable parameters, however, may be limiting the accuracy of the F12 methods relative to the other explicitly correlated analogs. All available evidence\cite{VRG:tew:2005:JCP,VRG:johnson:2017:CPL,VRG:may:2005:PCCP} suggests that the Slater-type geminal in \cref{eq:stg} is near optimal when used to correlate valence electron pairs. Another way to improve the accuracy would be to go back to the older ``orbital-invariant'' ansatz\cite{VRG:klopper:1991:CPL} which correlated each electron pair using an adjustable {\em linear} combination of $O^2$ (where $O$ is the number of occupied orbitals) explicitly correlated terms generated from a single fixed geminal; unfortunately this ansatz suffered from geminal superposition errors\cite{VRG:tew:2008:CPL} and poor conditioning of its optimization\cite{VRG:klopper:1999:CaL}.
Using multiple geminals\cite{VRG:valeev:2006:JCP} is another possibility but would suffer from the same issues as the orbital-invariant approach.

Fortunately, it turns out that there is still room for improvement of the standard F12 technology using single fixed Slater-type geminals. In the course of stretching the F12 technology to be usable with large (6Z and 7Z) correlation-consistent basis sets, we noticed that the optimal geminal lengthscale for the modern F12 formalism differs relatively strongly between low-order models like MP2-F12 and infinite-order models like CC-F12 (albeit with approximate inclusion of the F12 terms). Unfortunately, the recommended lengthscale parameters available in the literature were all determined at the MP2-F12 level of theory. Namely, Peterson and co-workers\cite{VRG:peterson:2008:JCP} determined the optimal geminal parameters by maximizing the magnitude of the MP2-F12 correlation energy of a collection of atoms and molecules spanning the 2nd and 3rd periods of the Periodic Table; they provided recommended geminal lengthscales for the augmented correlation-consistent basis sets\cite{VRG:dunning:1989:JCP,VRG:kendall:1992:JCP,VRG:woon:1993:JCP,VRG:woon:1994:JCP,VRG:wilson:1999:JCP} aug-cc-pV$X$Z (abbreviated as a$X$Z herein)  
for $X$ = D,T,Q,5 and the F12-specialized correlation consistent basis sets\cite{VRG:peterson:2008:JCP} cc-pV$X$Z-F12 (abbreviated as $X$Z-F12 herein) for $X$ = D,T,Q (see \cref{tbl:opt-beta}). Later, \citeauthor{VRG:hill:2009:JCP} performed a somewhat more robust optimization, again at the MP2-F12 level of theory, which resulted only in minor changes to the recommended exponents.\cite{VRG:hill:2009:JCP} The latter approach was also used to provide a recommended exponent for the 5Z-F12 basis \cite{VRG:peterson:2015:MP}. 
Although Hill et al. noticed that the optimal geminal exponents differed significantly between the MP2-F12 and CCSD-F12 methods, they attributed the difference to an artifact of the particular CCSD-F12 approximation (namely, the F12b approach\cite{VRG:adler:2007:JCP}) used in their work, in part because the CCSD-F12b energies obtained with the larger optimal exponents they found for F12b overshot their CBS limit estimate, which was obtained from CCSD-F12b computations with a large custom uncontracted basis. As additional evidence they referred to an earlier study 
by \citeauthor{VRG:tew:2007:PCCP} which concluded that the optimal geminal exponents agree well between MP2-F12 and another CCSD-F12 approximation (the CCSD(F12) method).\cite{VRG:tew:2007:PCCP} The study of Tew et al. used the older F12 formalism for explicitly-correlated MP2 and CCSD based on the non-diagonal orbital-invariant ansatz\cite{VRG:klopper:1991:CPL} rather than the more modern F12 formalism based on the SP ansatz of Ten-no\cite{VRG:ten-no:2004:JCP} (i.e., the diagonal F12 ansatz with amplitudes fixed by the exact spin-dependent cusp conditions, denoted with the ``(fix)''\cite{VRG:werner:2007:JCP} suffix by the Molpro team). Note that Knizia et al. had also observed the difference in the geminal exponent dependence of the SP-ansatz-based F12b and non-SP-ansatz (F12) coupled-cluster counterparts.\cite{VRG:knizia:2009:JCP}  Hill et al. did not associate the observed MP2 vs CC geminal lengthscale differences with the use of the diagonal ansatz, although some earlier evidence in a sufficiently different setting\cite{VRG:hofener:2009:CP} indicated that the diagonal F12 ansatz has a stronger dependence on the F12 geminal lengthscale than the nondiagonal counterpart, suggesting the diagonal and non-diagonal F12 results should be compared with caution.

In our own testing, we also observed a difference in the optimal geminal exponents between the MP2-F12 and CCSD-F12 methods using a perturbative approximation to the CCSD-F12 (namely, the CCSD$(2)_{\overline{\rm F12}}$ method\cite{VRG:valeev:2008:PCCP,VRG:torheyden:2008:PCCP}). Thus it became clear that it is necessary to reoptimize the geminal lengthscales for high-order F12 methods, such as the CC-F12 method as well as the transcorrelated F12 method.\cite{VRG:yanai:2012:JCP,VRG:masteran:2023:JCP} This manuscript reports our initial findings and the set of recommended geminal lengthscales for application with MP2-F12 and CC-F12 methods using the a$X$Z ($X=\text{D}\dots 6$) and $X$Z-F12 ($X=\text{D}\dots 5$) basis sets. Additionally, we provide a recommendation for the CC-F12 method with the a$7$Z basis.
In \cref{sec:technical} we describe the technical details of computations. \cref{sec:results} describes our protocol for optimizing $\beta$ and the analysis of the optimal values and their effect on the basis set incompleteness of absolute and relative correlation energies. In \cref{sec:conclusions} we summarize our findings.

\section{Technical Details}\label{sec:technical}

All the results reported herein were produced with a development version of the Massively Parallel Quantum Chemistry (MPQC) software package \cite{VRG:peng:2020:JCP}.
All correlation energies were obtained with only valence electrons correlated (frozen-core approximation). Both MP2 and CCSD F12 methods utilized the SP ansatz\cite{VRG:ten-no:2004:JCP}. The explicitly correlated CCSD energies were produced using the perturbative CCSD$(2)_{\overline{\mathrm{F12}}}$ approximation\cite{VRG:valeev:2008:PCCP, VRG:torheyden:2008:PCCP,VRG:valeev:2008:JCP,VRG:zhang:2012:JCTC} to the full CCSD-F12 method; for simplicity, we use CCSD-F12 to denote the former.
The F12 two-electron basis was generated from a single Slater-type geminal\cite{VRG:ten-no:2004:CPL} (\cref{eq:stg}), without the usual approximation as a linear combination of Gaussian Geminals.\cite{VRG:may:2004:JCP,VRG:tew:2005:JCP,VRG:may:2005:PCCP} 
Integrals of the STG and related integrals can be reduced to the core 1-dimensional integral,\cite{VRG:ten-no:2004:CPL,VRG:ahlrichs:2006:PCCPP}
\begin{align}
\label{eq:gm-tenno}
G_{m}(T,U) = \int_0^1 \mathrm{d}t \, t^{2m} \exp\left(-Tt^2 + U(1-t^{-2})\right), \quad m\leq-1
\end{align}
which in the Libint Gaussian AO integral engine are computed using 15th-order Chebyshev interpolation for $0 \leq T \leq 2^{10}$, $10^{-7} \leq U \leq 10^{3}$. For larger $T$ the upward recursion relation\cite{VRG:ten-no:2004:CPL} is used. Until this work, combinations of $T$ and $U$ outside of these ranges were not encountered.
However, the combination of low exponents present in high-$X$ OBS and higher $\beta$ than previously considered requires the evaluation of \cref{eq:gm-tenno} with $U \geq 10^3$ and small $T$.
To support evaluation of integrals of the STG and related integrals for extended range of Gaussian AO exponents and geminal parameters Libint version 2.11\cite{libint-2.11.0} introduced a new approach. Namely, outside of the ranges covered by the interpolation and upward recursion the STG is represented as a linear combination of Gaussian geminals using the approach developed in Refs. \citenum{VRG:harrison:2003:CS--I2,VRG:beylkin:2005:ACHA} and used in Ref. \citenum{VRG:bischoff:2014:JCP} with exponents and coefficients provided by trapezoidal quadrature of its integral representation
\begin{align}
-\frac{\exp(-\beta r) }{ \beta} = - \frac{2}{\sqrt{\pi}} \int_{-\infty}^{\infty} \mathrm{d}s \, \exp(s -\beta^2 \exp(2s) - r^2 \exp(-2 s)/4),
\end{align}
on interval $s\in[\log(\epsilon)/2-1,\log(T r_\text{lo}^{-2})/2]$ discretized evenly with step size $h=(0.2 - 0.5 \log_{10}(\epsilon))^{-1}$; $T=26$ is sufficient to ensure \{relative,absolute\} precision of $\epsilon=10^{-12}$ for $r$  between $r_\text{lo}=10^{-5}$ and \{$\beta^{-1}$,$\infty$\}, respectively.

Molecular orbitals were expanded in Dunning's aug-cc-pV$X$Z \cite{VRG:dunning:1989:JCP, VRG:kendall:1992:JCP,VRG:woon:1993:JCP,VRG:woon:1994:JCP,VRG:peterson:1994:JCP,VRG:wilson:1996:JMST,VRG:vanmourik:1999:MP,VRG:vanmourik:2000:IJQC,VRG:feller:1999:JCP,VRG:feller:2003:JCP} 
orbital basis sets (OBS), denoted a$X$Z, as well as the cc-pV$X$Z-F12 basis sets of Peterson and co-workers\cite{VRG:peterson:2008:JCP,VRG:peterson:2015:MP,VRG:sylvetsky:2016:JCP}, denoted $X$Z-F12. Robust density fitting in the aug-cc-pV\emph{X}Z-RI\cite{VRG:weigend:2002:JCP, VRG:hattig:2005:PCCP} density fitting basis set (DFBS) was used to approximate the 2-electron integrals throughout all computations. 3- and 4-electron integrals in the special F12 intermediates were approximated using the CABS+ approach\cite{VRG:valeev:2004:CPL} and approximation C.\cite{VRG:kedzuch:2005:IJQC} The aug-cc-pV$X$Z/OptRI\cite{VRG:yousaf:2009:CPL} and cc-pV$X$Z-F12/OptRI\cite{VRG:yousaf:2008:JCP} auxiliary basis sets (ABS) were used to approximate the F12 intermediates in computations with the a$X$Z and $X$Z-F12 OBS, respectively. a$X$Z/OptRI, $X$Z-F12/OptRI, and a$X$Z-RI basis sets are only available for $X\leq 5$, $X\leq$ Q, $X\leq 6$, respectively, hence for computations with higher $X$ we used the largest respective basis that is available; e.g.,  a6Z and a7Z OBSs were matched by the a5Z/OptRI ABS. 
The only exception were computations with 5Z-F12 OBS; the a5Z/OptRI basis set was used instead of QZ-F12/OptRI.
Significant errors were observed when extrapolating energies with the a7Z results computed with the a6Z-RI basis set. Thus, we used an automatically generated density fitting basis set using the MADF approach described in Ref. \citenum{surjuse2025physicsdrivenconstructioncompactprimitive}. 

All Gaussian AO basis sets not already included in the Libint library\cite{libint-2.11.0} were obtained from the Basis Set Exchange\cite{VRG:schuchardt:2007:JCIM,VRG:pritchard:2019:JCIM} except the a7Z basis for hydrogen, which was provided by John F. Stanton's research group.

The CBS CCSD valence correlation energies and their contributions to the atomization energies were obtained by the $X^{-3}$ extrapolation\cite{VRG:helgaker:1997:JCP} of the a6Z and a7Z energies. The corresponding CCSD contributions to the reaction energies and ionization potentials (IPs) used $X^{-3}$ extrapolation from the aQZ and a5Z energies. The ``Silver'' benchmark CCSD values from \citeauthor{VRG:rezac:2011:JCTC}'s original paper were used as the CCSD reference binding energies for the S66 benchmark. \cite{VRG:rezac:2011:JCTC}

\section{Results}\label{sec:results}

\subsection{Geminal lengthscale optimization}
\label{sec:results-beta}

The optimal inverse lengthscale $\beta_\text{opt}$ for the given combination of OBS and F12 method was determined by minimizing the following objective function:
\begin{align}
\label{eq:Ebarβ}
  \bar{E}_{\text{F12}}(\beta) = \left\langle \frac{E_{\text{F12}}^S(\beta)}{E_{\text{F12}}^S(\beta^S_\text{opt})} \right\rangle
\end{align}
where $E_{\text{F12}}^S(\beta)$ is the F12 contribution to the energy of system $S$, $\beta^S_\text{opt}$ is the value of $\beta$ that minimizes $E_{\text{F12}}^S(\beta)$, and $\langle \dots \rangle$ denotes the averaging over $S$.
The purpose of the denominator in \cref{eq:Ebarβ} is to renormalize each fit according to the energy of each system to avoid giving excessive weight to the systems with large correlation energies, thereby balancing the correlation physics of across the periods and groups of the Periodic Table.

To reduce the cost of optimization, $E_{\text{F12}}^S(\beta)$ for each system was approximated by a quartic polynomial obtained by least-squares fitting $E_{\text{F12}}^S(\beta)$ evaluated on an equidistant grid of $\beta$ values (use of 6th-order polynomial changed the optimal exponents by 0.01 or so). For each basis/method combination the grid was a set of points $\beta_i \equiv i/20, i\in \mathbb{Z}$ selected such that at least 3 grid points were included on each side of $\beta^S_\text{opt}$ for every $S$ in the training set. The training set of systems $\{S\}$ included lowest singlet states of dimers $\text{A}_2$ and hydrides $\text{AH}_x$, with A including elements from groups 13-17 in the 2nd and 3rd period of the Periodic Table, as well as the \ce{Ne} and \ce{Ar} atoms. Namely, the benchmark set consists of \ce{B2}, \ce{BH3}, \ce{C2}, \ce{CH4}, \ce{N2}, \ce{NH3}, \ce{O2}, \ce{H2O}, \ce{F2}, \ce{HF}, \ce{Ne}, \ce{Al2}, \ce{AlH3}, \ce{Si2}, \ce{SiH4}, \ce{P2}, \ce{PH3}, \ce{S2}, \ce{H2S}, \ce{Cl2}, \ce{HCl}, and \ce{Ar}.
All computations were performed at the equilibrium CCSD(T)/aTZ geometries obtained from CCCBDB \cite{VRG:johnson:2002:}. 
Since the a7Z basis is not available for the 3rd period elements, only the molecules composed of 2nd period elements were used for the geminal optimization with that basis. Additionally, treating \ce{CH4} with the a7Z basis proved too challenging due to noisy DF errors with the recently designed MADF basis, so it was omitted from the optimization set in this case.

\begin{table}[htb]
    \centering
    \begin{tabular}{l|c|c|c|c|c|c|c|c} \hline \hline
        OBS Family & Source & Training Method & D & T & Q & 5 & 6 & 7 \\
        \hline 
       \multirow{4}{*}{aXZ} & Ref. \citenum{VRG:peterson:2008:JCP} & MP2-F12 & 1.1 & 1.2 & 1.4 & 1.4 & -- & -- \\
       & Ref. \citenum{VRG:hill:2009:JCP} & MP2-F12 & 1.0 & 1.2 & 1.4 & 1.5 & -- & -- \\
       & \textbf{this work} & \textbf{MP2-F12} & \textbf{0.96} & \textbf{1.21} & \textbf{1.46} & \textbf{1.49} & \textbf{1.51} & -- \\
       & \textbf{this work} & \textbf{CCSD-F12 }& \textbf{1.12 }& \textbf{1.61 }& \textbf{2.16 }& \textbf{2.57} & \textbf{2.98} & \textbf{3.85}  \\ \hline
       \multirow{4}{*}{XZ-F12} & Ref. \citenum{VRG:peterson:2008:JCP} & MP2-F12 & 0.9 & 1.0 & 1.1 & -- & -- & -- \\
       & Ref. \citenum{VRG:hill:2009:JCP} & MP2-F12 & 0.9 & 1.0 & 1.0 & 1.2$^a$ & -- & -- \\
       & \textbf{this work}  & \textbf{MP2-F12} & \textbf{0.84} & \textbf{0.93} & \textbf{0.92} & \textbf{1.02} & \textbf{--} & \textbf{--} \\
       &  \textbf{this work} & \textbf{CCSD-F12} & \textbf{1.06 }& \textbf{1.52 }& \textbf{1.95 }& \textbf{2.31} & \textbf{--} & \textbf{--} \\       
     \hline\hline
    \end{tabular}
    \caption{Optimal $\beta$ values recommended in this work.}
    \label{tbl:opt-beta}
    $^a$ Obtained in Ref. \citenum{VRG:peterson:2015:MP} using optimization method of Ref. \citenum{VRG:hill:2009:JCP}.
\end{table}

The resulting optimal geminal parameters $\beta_\text{opt}$ 
as well as the recommended values from Refs. \citenum{VRG:peterson:2008:JCP,VRG:hill:2009:JCP,VRG:peterson:2015:MP} are listed in \cref{tbl:opt-beta}.

There is excellent agreement between our MP2-F12 $\beta_\text{opt}$ and those obtained previously by Peterson and co-workers, despite the differences in the training sets and technical details; most deviations are smaller than 0.1. The largest deviation of 0.18 is observed for the cc-pV5Z-F12 basis; as we will see shortly, tolerance of small errors in $\beta_\text{opt}$ rapidly increases with $X$ due to the rapid decrease of the curvature of $\bar{E}_{\text{F12}}(\beta)$ near the minimum.

\begin{figure}[ht!]
    \centering
    \includegraphics[width=0.5\linewidth]{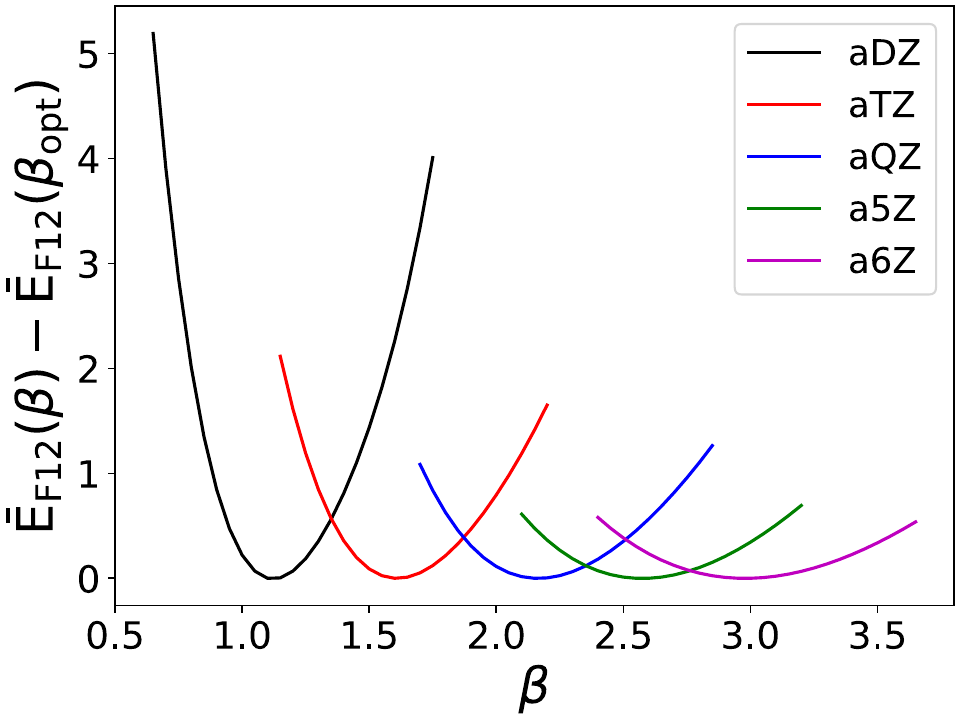}
    \caption{Levelized CCSD-F12 $\bar{E}_{\text{F12}}(\beta)$ for the a$X$Z OBS family. Optimal $\beta$ increases with the cardinal number $X$, whereas the curvature of $\bar{E}_{\text{F12}}(\beta)$ decreases with $X$.}
    \label{fig:all-axz-fits}
\end{figure}

The most notable insight from \cref{tbl:opt-beta} is the rapid increase of the gap between CCSD-F12 and MP2-F12 $\beta_\text{opt}$; while for the double-zeta basis sets the gap is modest ($\approx 0.2$) it exceeds 1 for quintuple zeta basis sets and aparently continues to grow thereafter.
Note that the $\beta_\text{opt}$ obtained by \citeauthor{VRG:hill:2009:JCP} using CCSD-F12b with the a5Z basis, 2.4 $a^{-1}_0$,\cite{VRG:hill:2009:JCP} is in good agreement with our $\beta_\text{opt}$ value, 2.57 $a^{-1}_0$, despite the substantial differences between the the iterative CCSD-F12b approximation and perturbative CCSD$(2)_{\overline{\rm F12}}$ approximation to exact CCSD-F12.
Plots of $\bar{E}_{\text{F12}}(\beta)$ in \cref{fig:all-axz-fits} illustrate the consequences of using MP2-F12-optimized exponents for large OBS: although the dependence on $\beta$ weakens with $X$ and thus high-$X$ F12 energies are more tolerant of small errors in $\beta_\text{opt}$, the CCSD-MP2 gap $\beta_\text{opt}$ is large and grows with $X$, hence using MP2-F12-optimized $\beta_\text{opt}$ will result in suboptimal CCSD-F12 energies for $X\geq3$.

\Cref{tbl:bsie} illustrates quantitatively the dependence of CCSD-F12 basis set incompleteness error (BSIE) on the geminal lengthscale. The fifth column illustrates the reduction of the CCSD-F12 BSIE by switching from the original geminal lengthscales of \citeauthor{VRG:peterson:2008:JCP} to our per-basis optimal CCSD-F12 parameters listed in \cref{tbl:opt-beta}; the BSIEs are reduced by more than 60\% for quadruple-zeta basis sets and by more than a factor of 2 for quintuple-zeta basis sets. The use of CCSD-F12 rather than MP2-F12 optimized geminal parameters for yet larger OBS results in progressively larger BSIE reductions, by a factor of 4 for a7Z! Clearly, the use of CCSD-F12-optimized geminal parameters for CCSD-F12 and other infinite-order F12 methods should be preferred.

\begin{table}[ht!]
\begin{tabular}{lrrrrrr} \hline \hline
OBS & $\langle \delta (\beta_\text{ref}) \rangle$ & $\langle \delta (\beta_\text{opt}) \rangle$ & $\langle \delta (\beta^S_\text{opt}) \rangle$ & $\langle \delta (\beta_\text{ref}) / \delta (\beta_\text{opt}) \rangle$ & $\langle \delta (\beta_\text{ref}) / \delta (\beta^S_\text{opt}) \rangle$ & $\langle \delta (\beta_\text{opt}) / \delta (\beta^S_\text{opt}) \rangle$ \\ \hline
aDZ & 17.42 & 17.30 & 16.02 & 1.005 & 1.095 & 1.087 \\
aTZ & 7.76 & 6.11 & 5.91 & 1.244 & 1.287 & 1.032 \\
aQZ & 3.48 & 2.10 & 2.04 & 1.646 & 1.691 & 1.030 \\
a5Z & 1.71 & 0.85 & 0.83 & 2.187 & 2.235 & 1.024 \\
a6Z & 1.11 & 0.45 & 0.44 & 2.513 & 2.584 & 1.029 \\ 
a7Z & 0.36 & 0.08 & 0.08 & 4.855 & 4.899 & 1.008 \\ \hline
DZ-F12 & 13.05 & 12.04 & 11.29 & 1.082 & 1.153 & 1.060 \\
TZ-F12 & 5.07 & 3.86 & 3.75 & 1.414 & 1.453 & 1.028 \\
QZ-F12 & 1.95 & 1.20 & 1.17 & 1.913 & 1.958 & 1.023 \\
5Z-F12 & 1.03 & 0.51 & 0.50 & 2.592 & 2.651 & 1.020 \\ \hline\hline
\end{tabular}
\caption{Average basis set incompleteness errors of CCSD-F12 correlation energies (m$E_\text{h}$) obtained with the recommended geminal parameters of Ref. \citenum{VRG:peterson:2008:JCP,VRG:peterson:2015:MP} ($\beta_\text{ref}$; $\beta_\text{ref}=1.4$ was used for a6Z and a7Z MP2-F12), our per-basis recommended CCSD-F12 parameters from \cref{tbl:opt-beta} ($\beta_\text{opt}$) and their system-optimized counterparts ($\beta_\text{opt}^S$).$^a$}
\label{tbl:bsie}
$^a$ $\delta(\beta) \equiv E^\text{CCSD-F12}_\text{corr}(\beta) - E^\text{CBS}_\text{corr}$.
\end{table}

\begin{figure}[ht!]
    \centering
     \begin{subfigure}{0.49\textwidth}
    \includegraphics[width=0.95\linewidth]{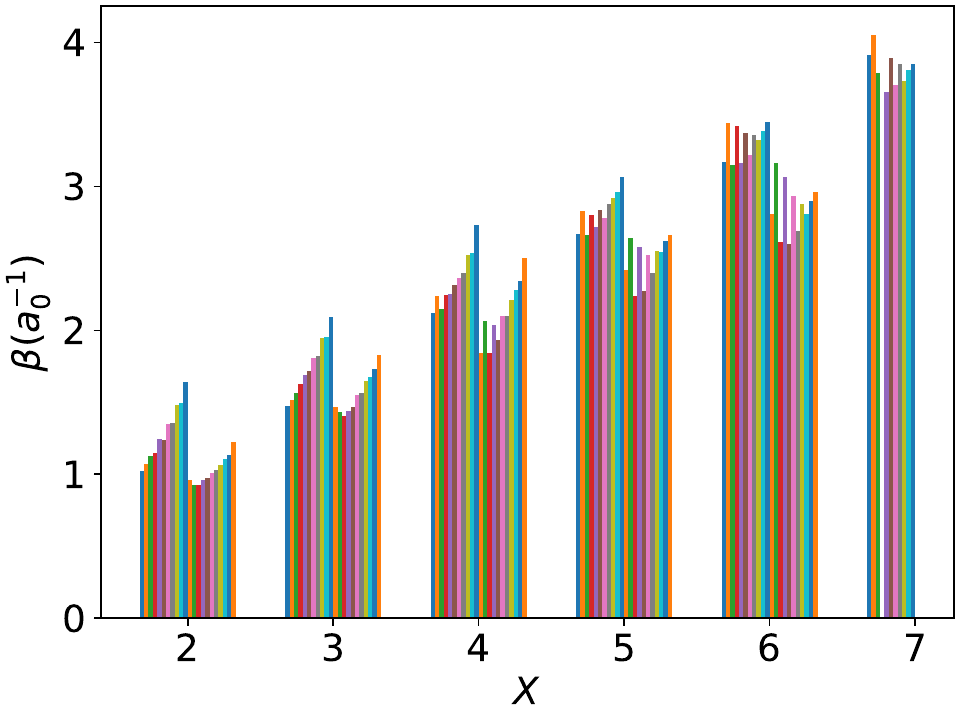}
    \caption{a$X$Z $\beta^{S}_\text{opt}$.}
    \label{fig:beta-opt-axz-distribution}
    \end{subfigure} 
     \begin{subfigure}{0.49\textwidth}
    \includegraphics[width=0.95\linewidth]{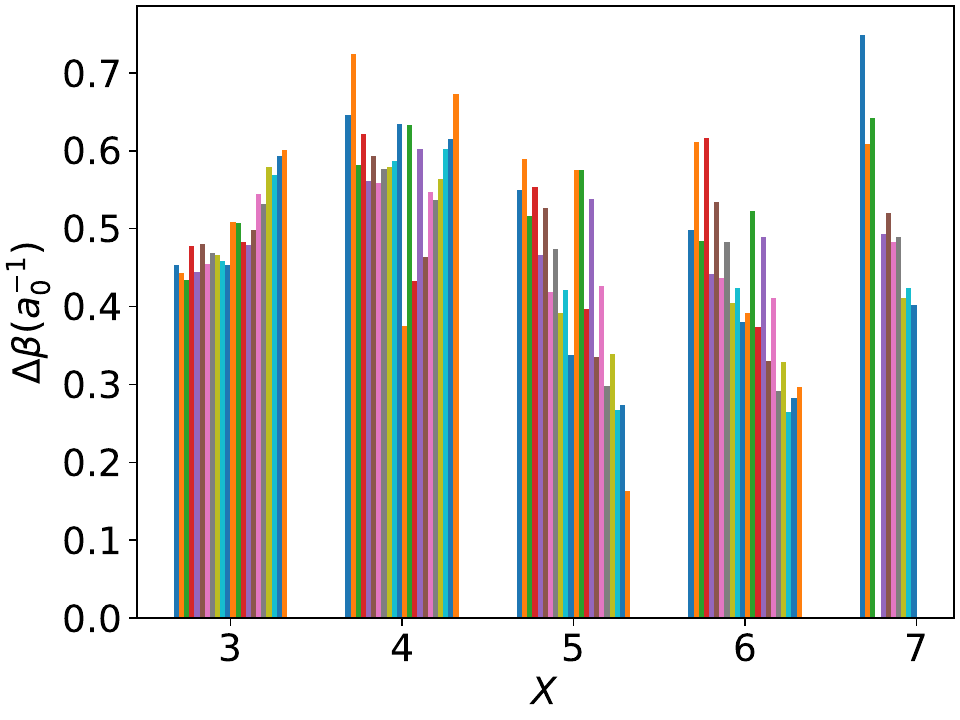}
    \caption{a$X$Z $\beta^{S}_\text{opt}$ - a$(X-1)$Z $\beta^{S}_\text{opt}$.}
    \label{fig:delta-beta-opt-axz-distribution}
    \end{subfigure} \\
     \begin{subfigure}{0.49\textwidth}
    \includegraphics[width=0.95\linewidth]{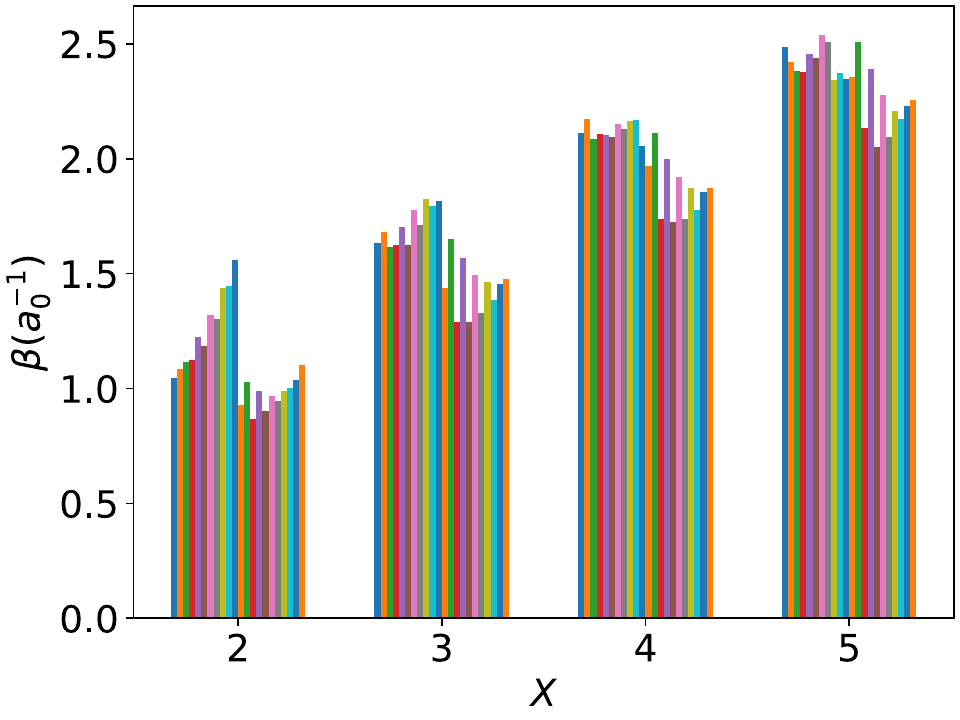}
    \caption{$X$Z-F12 $\beta^{S}_\text{opt}$.}
    \label{fig:beta-opt-xzf12-distribution}
    \end{subfigure}
     \begin{subfigure}{0.49\textwidth}
    \includegraphics[width=0.95\linewidth]{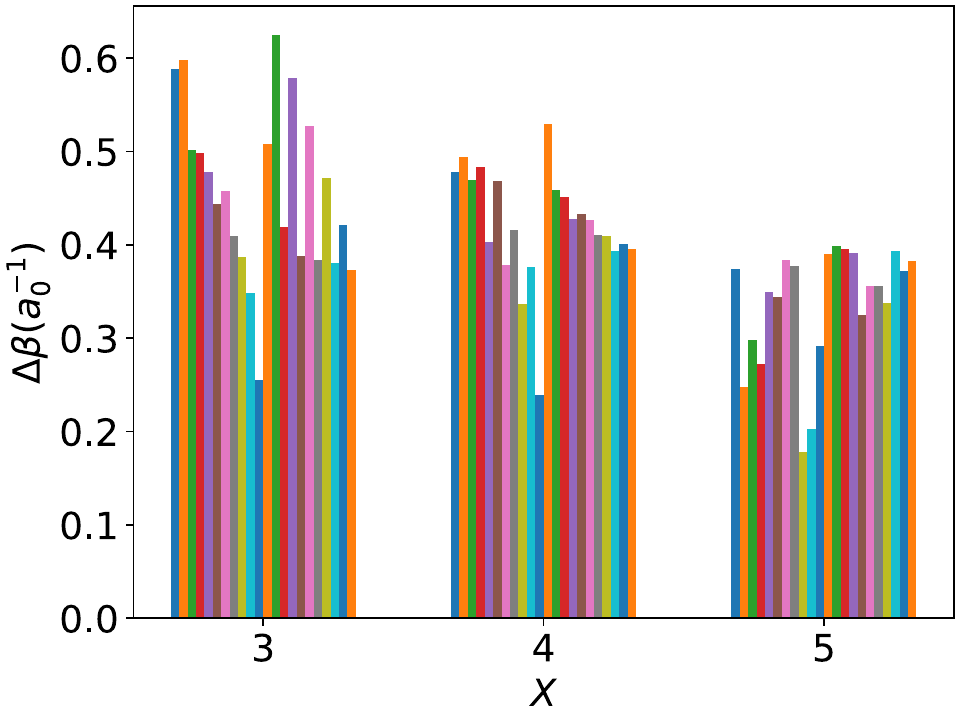}
    \caption{$X$Z-F12 $\beta^{S}_\text{opt}$ - $(X-1)$Z-F12 $\beta^{S}_\text{opt}$.}
    \label{fig:delta-beta-opt-xzf12-distribution}
    \end{subfigure}
    \caption{Distribution of CCSD-F12-optimized geminal parameters $\beta^S_\text{opt}$ across the training set. The bars represent each system in the training set with the same order of systems as listed in \cref{sec:results-beta}.}
    \label{fig:beta-opt-distribution}
\end{figure}

Note that the optimal geminal parameters vary substantially across the period and group, even for valence electrons.
\Cref{fig:beta-opt-axz-distribution,fig:beta-opt-xzf12-distribution} illustrate the distribution of $\beta_\text{opt}^S$ across the training set $\{S\}$. Several trends are noticeable. First, $\beta_\text{opt}^S$ varies most strongly
across the second period, and less strongly across the third period. Second, variation across the period weakens with increasing $X$. Third, $\beta_\text{opt}^S$ for the third period elements is nearly universally smaller than that for the second period. Despite the noticeable variation of $\beta_\text{opt}^S$ across the period and group of the Periodic Table, system-specific optimization of $\beta$ is results in a relatively small benefit compared to the use of per-basis $\beta_\text{opt}$, as illustrated by the last column of \cref{tbl:opt-beta}. Namely, the BSIE is only reduced by $\approx 3\%$ by system-specific optimization. Considering that the use of basis-tuned rather than system-tuned geminal parameters greatly simplifies the practical use of F12 methods (by, e.g., preserving their size consistency), in our opinion system-specific geminal optimization in the context of F12 methods is not worthwhile.

\begin{figure}
    \centering
    \includegraphics[width=\linewidth]{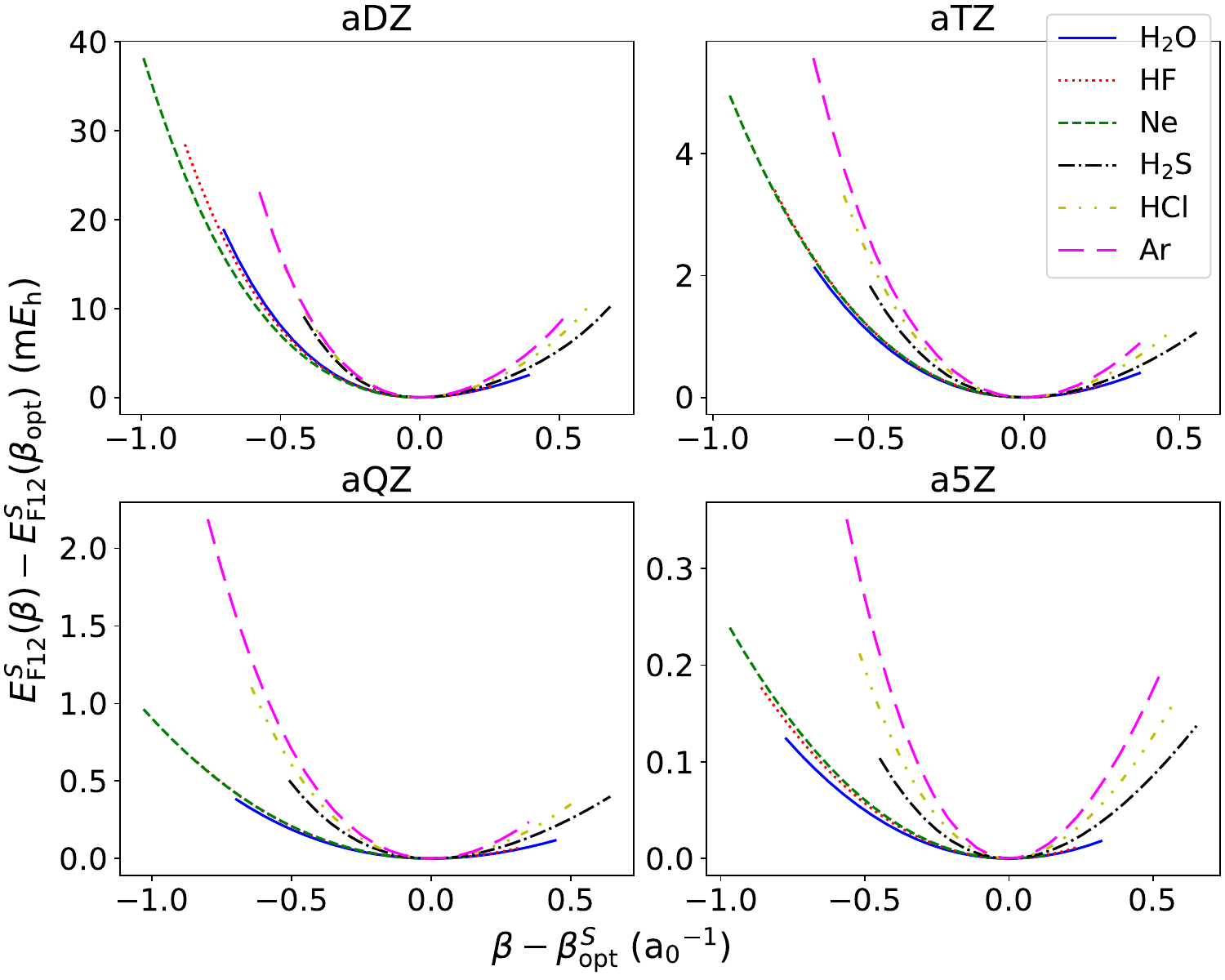}
    \caption{$E^S_\text{F12}(\beta) - E^S_\text{F12}(\beta^S_\text{opt})$ for several representative valence-isoelectronic systems with second- and third-period elements. Note the relatively weak variation of the curvature across the period, but substantial curvature increase upon transition from second to third period.}
    \label{fig:molfit-comparison}
\end{figure}

Variation with the period and group is observed not only for the position of the minimum of $E^S_\text{F12}(\beta)$ (i.e., $\beta^S_\text{opt}$) but also its curvature, as illustrated by \cref{fig:molfit-comparison}. It is somewhat unexpected that for heavier elements the F12 energy is more strongly dependent on $\beta$.

\Cref{fig:delta-beta-opt-axz-distribution,fig:delta-beta-opt-xzf12-distribution} illustrate that $\beta_\text{opt}^S$ varies relatively regularly with $X$, although the distribution changes shape and the noticeable differences between dimers and hydrides become pronounced for large $X$.

The strong increase of $\beta_\text{opt}$ upon transition from MP2-F12 to CCSD-F12 must correlate to the wider basis set error of the Coulomb hole of the MP1 wave function compared to its CCSD counterpart. Our findings agree with other observations regarding the differences between MP2 and infinite-order methods. For example, it is known that the basis set errors of MP2 are generally larger than those of higher-order methods like CCSD (e.g., see the relevant discussion in Ref. \citenum{VRG:valeev:2008:PCCP}). Also note the substantial differences in the optimal unoccupied orbital basis for MP2 vs CCSD studied in Ref. \citenum{VRG:clement:2018:JCTC}.

\subsection{Effect of Geminal Optimization on BSIEs of Relative Energies} \label{sec:results:properties}

Does the observed reduction in the BSIEs of {\em absolute} CCSD-F12 energies translate into reduced BSIEs of {\em relative} energies? To answer this question we report the CCSD correlation BSIEs for a variety of relative energies obtained with $\beta_\text{ref}$ and $\beta_\text{opt}$ in \cref{tbl:properties}. These include (1) a set of 15 reaction energies (see Table 9 in Ref. \citenum{VRG:tew:2007:PCCP}) involving small molecules with second- and third-period elements, (2) a set of 5 binding energies of weakly-bound pairs of molecules selected from the S66 benchmark set (namely water dimer, ethyne dimer, methane dimer, water-methane pair, and ethyne-water pair) \cite{VRG:rezac:2011:JCTC}, (3) ionization potentials (IPs) of homonuclear second-period diatomics (\ce{B2}, \ce{C2}, \ce{N2}, \ce{O2}, \ce{F2}, at their experimentally-derived equilibrium geometries\cite{VRG:johnson:2002:}) and (4) atomization energies of several molecules from the HEAT dataset (\ce{N2}, \ce{H2O}, \ce{HF}, \ce{F2}, \ce{OH}, \ce{CH}, \ce{NH3}, at geometries revised by John F. Stanton's research group and listed in Supporting Information).

\begin{table}[htbp]
\begin{tabular}{rrrrrrrrr}
\hline \hline
& \multicolumn{2}{c}{aDZ} & \multicolumn{2}{c}{aTZ} & \multicolumn{2}{c}{aQZ} & \multicolumn{2}{c}{a5Z} \\ 
 & $\beta_\text{ref}$ & $\beta_\text{opt}$ & $\beta_\text{ref}$ & $\beta_\text{opt}$ & $\beta_\text{ref}$ & $\beta_\text{opt}$ & $\beta_\text{ref}$ & $\beta_\text{opt}$ \\ \hline
 & \multicolumn{8}{c}{reaction energies} \\ 
$\langle\delta\rangle$ & 0.666 & 0.657 & 0.359 & 0.489 & -0.167 & -0.078 & -0.064 & -0.003 \\
$\langle|\delta|\rangle$ & 1.915 & 1.920 & 0.958 & 1.343 & 0.624 & 0.692 & 0.404 & 0.421 \\
$\sigma$ & 2.570 & 2.594 & 1.474 & 1.940 & 0.791 & 0.997 & 0.535 & 0.617 \\
& \multicolumn{8}{c}{noncovalent interaction energies} \\
$\langle\delta\rangle$ & -0.478 & -0.477 & -0.353 & -0.394 & -0.202 & -0.214 & & \\
$\langle|\delta|\rangle$ & 0.478 & 0.477 & 0.353 & 0.394 & 0.202 & 0.214 & & \\
$\sigma$ & 0.109 & 0.111 & 0.097 & 0.076 & 0.051 & 0.054 & & \\ 
& \multicolumn{8}{c}{ionization potentials} \\
$\langle\delta\rangle$ & 95.6 & 95.3 & 24.8 & 24.1 & 5.9 & 6.2 & 3.1 & 2.4 \\
$\langle|\delta|\rangle$ & 95.6 & 95.3 & 24.8 & 24.1 & 5.9 & 6.2 & 3.1 & 2.4 \\
$\sigma$ & 59.6 & 58.6 & 20.4 & 18.9 & 5.1 & 4.4 & 1.6 & 1.5 \\ 
& \multicolumn{8}{c}{atomization energies} \\
$\langle\delta\rangle$ & 4.535 & 4.381 & 1.962 & 1.608 & -0.271 & -0.319 & 0.027 & -0.132 \\%& 0.046 & -0.180 & -0.002 & -0.158 \\
$\langle|\delta|\rangle$ & 6.154 & 6.052 & 2.066 & 1.864 & 0.433 & 0.399 & 0.140 & 0.135 \\ %& 0.066 & 0.180 & 0.074 & 0.158 \\
$\sigma$ & 5.567 & 5.551 & 1.609 & 1.498 & 0.523 & 0.453 & 0.182 & 0.129 \\
\hline \hline %& 0.109 & 0.180 & 0.103 & 0.158
 & \multicolumn{2}{c}{DZ-F12} & \multicolumn{2}{c}{TZ-F12} & \multicolumn{2}{c}{QZ-F12} & \multicolumn{2}{c}{5Z-F12}  \\ 
& $\beta_\text{ref}$ & $\beta_\text{opt}$ & $\beta_\text{ref}$ & $\beta_\text{opt}$ & $\beta_\text{ref}$ & $\beta_\text{opt}$ & $\beta_\text{ref}$ & $\beta_\text{opt}$ \\ \hline
& \multicolumn{8}{c}{reaction energies} \\ 
$\langle\delta\rangle$ & 0.508 & 0.493 & 0.186 & 0.258 & 0.096 & 0.098 & & \\
$\langle|\delta|\rangle$ & 1.415 & 1.591 & 0.594 & 1.049 & 0.467 & 0.566 & & \\
$\sigma$ & 1.723 & 1.832 & 0.845 & 1.472 & 0.692 & 0.853 & & \\ 
& \multicolumn{8}{c}{noncovalent interaction energies} \\
$\langle\delta\rangle$ & -0.118 & -0.192 & -0.167 & -0.159 & -0.063 & -0.067 \\
$\langle|\delta|\rangle$ & 0.118 & 0.192 & 0.167 & 0.159 & 0.063 & 0.067 \\
$\sigma$ & 0.033 & 0.083 & 0.052 & 0.075 & 0.027 & 0.038 \\
& \multicolumn{8}{c}{ionization potentials} \\
$\langle\delta\rangle$ & 76.1 & 68.9 & 23.0 & 19.0 & 8.7 & 5.6 & 4.1 & 2.0 \\
$\langle|\delta|\rangle$ & 76.1 & 68.9 & 23.0 & 19.0 & 8.7 & 5.6 & 4.1 & 2.0 \\
$\sigma$ & 44.8 & 38.5 & 14.6 & 13.2 & 4.8 & 4.3 & 1.6 & 1.7 \\ 
& \multicolumn{8}{c}{atomization energies} \\
$\langle\delta\rangle$ & 7.621 & 5.938 & 2.409 & 1.516 & 0.688 & 0.170 & 0.259 & -0.079 \\
$\langle|\delta|\rangle$ & 7.621 & 5.938 & 2.409 & 1.520 & 0.688 & 0.213 & 0.259 & 0.095 \\
$\sigma$ & 3.347 & 2.996 & 1.268 & 1.000 & 0.378 & 0.173 & 0.143 & 0.151 \\ \hline\hline
\end{tabular}
\caption{Statistical analysis of the BSIEs of CCSD correlation energy contributions to various relative energies.$^a$}
\label{tbl:properties}
$^a$ $\langle\delta\rangle$,  $\langle|\delta|\rangle$, and $\sigma$ denote mean signed, mean unsigned, and standard deviation of BSIE, $\delta \equiv E - E_\text{CBS}$. IP BSIEs are in meV, the rest of the values are kJ/mol.
\end{table}

As expected, relative CCSD-F12 correlation energies are less affected by the optimization of $\beta$ than absolute energies. 
The BSIEs of reaction energies, in fact, generally appear to be marginally worse with $\beta_\text{opt}$ than $\beta_\text{ref}$. The BSIEs of noncovalent interaction energies are only marginally affected by the changes in $\beta$. Ionization potentials and atomization energies are the only properties where significant improvement is observed for some basis sets. The most significant differences are observed for the $X$Z-F12 basis sets, but for the largest a$X$Z basis sets some improvement is also observed. E.g., even for the \{T,Q\}Z-F12 OBS the BSIEs of CCSD-F12 correlation atomization energies are reduced from \{2.41, 0.69\} to \{1.52, 0.21\} kJ/mol, i.e. by \{37, 61\} \%. 
As expected, CCSD-F12 computations for the largest basis sets benefit most from the use of $\beta_\text{opt}$.

\begin{figure}[ht!]
    \centering
    \includegraphics[width=\linewidth]{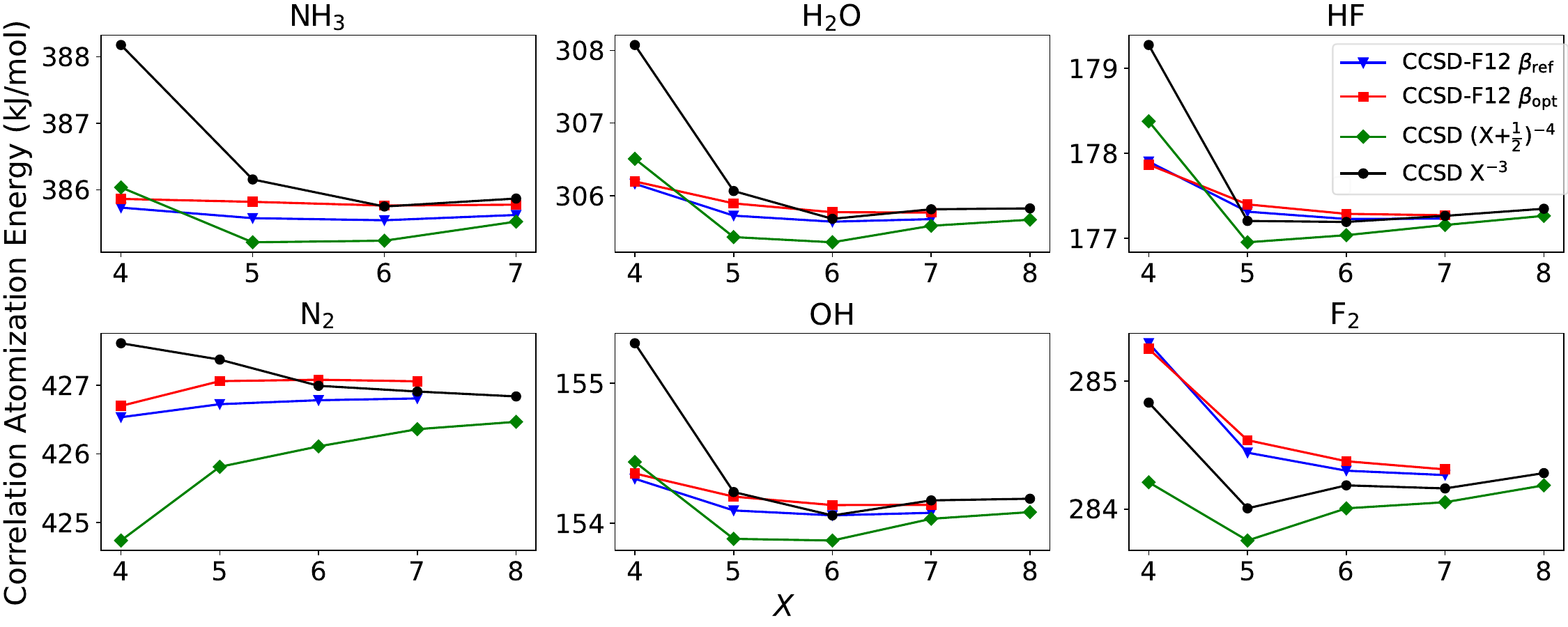}
    \caption{Basis set convergence of a$X$Z CCSD-F12 and a$\{X-1,X\}$Z extrapolated CCSD correlation contributions to atomization energies of several small molecules in the HEAT benchmark set.\cite{VRG:tajti:2004:JCP} $\beta_\text{ref}$ and $\beta_\text{opt}$ denote the use of geminal parameters from Ref. \citenum{VRG:peterson:2008:JCP} (extended to use $\beta_\text{ref}=1.4$ for a6Z OBS) and \cref{tbl:opt-beta}, respectively.}
    \label{fig:atomization-extrap}
\end{figure}

The original motivation for this work was to reduce the BSIEs of coupled-cluster energies in accurate thermochemical benchmarks, such as HEAT,\cite{VRG:tajti:2004:JCP,VRG:harding:2008:JCP} by replacing the high-end extrapolation of CCSD(T) energies with extrapolation-free explicitly correlated CCSD(T) energies. Optimization of the geminal parameters with the largest a6Z 
basis sets allows us for the first time to probe whether the F12 methods are competitive with extrapolation in this regime. \Cref{fig:atomization-extrap} illustrates the CCSD correlation contribution to atomization energies obtained with extrapolation and with CCSD-F12. Indeed, it appears that the F12 energies should be preferred to extrapolation, as the basis set convergence of the former appears more systematic and rapid than that of the latter. The optimization of $\beta$ does not appear to have a significant effect on convergence, which is in agreement with the nearly identical performance of $\beta_\text{ref}$ and $\beta_\text{opt}$ a$X$Z CCSD-F12 for atomization energies in \cref{tbl:properties}. According to the data in \cref{tbl:properties} we expect the convergence of $X$Z-F12 atomization energies to be substantially improved by the use of $\beta_\text{opt}$. In any case, the BSIEs of the F12 energies seems to be significantly reduced relative to that of the extrapolated counterparts; the average unsigned ($X=6$) -- ($X=5$) difference for the F12 $\beta_\text{opt}$ energies is 0.077 kJ/mol whereas the corresponding value for the a$\{X-1,X\}$Z $X^{-3}$-extrapolated energies is 0.244 kJ/mol.

Although relative CCSD-F12 correlation energies are not as significantly affected by the reoptimization of $\beta$ as are the absolute counterparts, in some important regimes the use of CCSD-F12-optimized $\beta$ yield major reduction of BSIE. Thus the use of the new CCSD-F12-optimized $\beta$ values is recommended for F12 variants of all accurate (i.e., infinite-order) model chemistries, such as the CC-F12 methods.

\section{Conclusions}\label{sec:conclusions}
We have reported updated recommendations for the geminal lengthscale parameter for F12 calculations, showing that the previous values optimized using the MP2-F12 method\cite{VRG:peterson:2008:JCP,VRG:hill:2009:JCP} are suboptimal for higher-order F12 methods formulated using the SP 
 (diagonal fixed-coefficient spin-adapted) F12 ansatz. 
The new geminal parameters are shown to reduce the BSIEs of absolute valence CCSD-F12 correlation energies, by a significant (and increasing with the cardinal number of the basis) margin. The effect of geminal reoptimization is especially pronounced for the cc-pV$X$Z-F12 basis sets (specifically designed for use with F12 methods) relative to their conventional aug-cc-pV$X$Z counterparts. The BSIEs of relative energies are less affected but substantial reductions can be obtained, especially for atomization energies and ionization potentials with the cc-pV$X$Z-F12 basis sets.
The new geminal parameters are therefore recommended for all applications of coupled-cluster F12 methods.

It remains to be seen how strongly the optimal geminal parameters depend on the form in which F12 terms are included and on the level of correlation treatment. In limited testing only negligible differences were found in the geminal exponents optimized with CCSD-F12 and CCSDT-F12 methods. Slightly larger differences were observed between the optimal exponents between traditional incorporation of F12 terms into the cluster operator of CC-F12 and their a priori introduction via the F12-style transcorrelation.\cite{VRG:yanai:2012:JCP,VRG:masteran:2023:JCP,VRG:ten-no:2023:JCP}
A more thorough investigation of these effects is underway and will be reported elsewhere.

\begin{suppinfo}

Molecular geometries, select raw computational data.

\end{suppinfo}

\begin{acknowledgement}
We gratefully acknowledge late John F. Stanton (University of Florida) for helping kickstart this research project; his neverending quest for pushing molecular many-body methods to match and beat the experimental accuracy prompted us to push the F12 explicit correlation technology to its limits.

This research was supported by the US Department of Energy, Office of Science, via award DE-SC0022327. The development of the \code{Libint} software library is supported by the Office of Advanced Cyberinfrastructure, US National Science Foundation via award 2103738. The development of the \code{SeQuant} software library is supported by the US National Science Foundation via award 2217081. The authors acknowledge Advanced Research Computing at Virginia Tech (\url{https://arc.vt.edu/}) for providing computational resources and technical support that have contributed to the results reported within this paper.
\end{acknowledgement}

\bibliography{vrgrefs, misc, srprefs} 

\newpage
\thispagestyle{empty}

{\bf For Table Of Contents Only} \\

\centering
\includegraphics[]{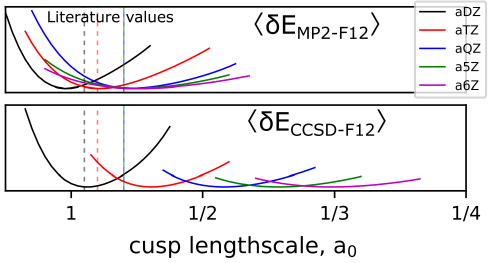} 
    
\end{document}

\maketitle

\section{Molecular Geometries}
Below are molecular geometries for the 7 molecules included in the atomization set. These are HEAT geometries (in atomic units) revised by John F. Stanton's research group.

\begin{verbatim}
**** CH ****
2

C -0.00000000     0.00000000     0.16372625
H  0.00000000     0.00000000    -1.94946043
**** F2 ****
2

F -0.00000000     0.00000000     1.33445279
F  0.00000000     0.00000000    -1.33445279
**** H2O ****
3

H  0.00000000    -1.43108710     0.98391550
O  0.00000000    -0.00000000    -0.12399124
H  0.00000000     1.43108710     0.98391550
**** HF ****
2

H -0.00000000     0.00000000     1.64546855
F  0.00000000     0.00000000    -0.08728862
**** N2 ****
2

N -0.00000000     0.00000000     1.03701651
N  0.00000000     0.00000000    -1.03701651
**** NH3 ****
4

N   -0.12753001     0.00000000     0.0000000
H    0.59064885    -0.88517861    -1.5331743
H    0.59064885     1.77035723    -0.0000000
H    0.59064885    -0.88517861     1.5331743
**** OH ****
2

O -0.00000000     0.00000000     0.10862251
H  0.00000000     0.00000000    -1.72391805
****
\end{verbatim}

\section{Select Raw Data}
\begin{table}[]
\begin{tabular}{rrrrrrrrrr}
\hline
 & \multicolumn{2}{c}{aDZ}  & \multicolumn{2}{c}{aTZ} & \multicolumn{2}{c}{aQZ} & \multicolumn{2}{c}{a5Z} & CBS \\
Rxn. & $\beta_\text{ref}$ & $\beta_\text{opt}$ & $\beta_\text{ref}$ & $\beta_\text{opt}$ & $\beta_\text{ref}$ & $\beta_\text{opt}$ & $\beta_\text{ref}$ & $\beta_\text{opt}$ & \\ \hline
1 & -26.4240 & -26.4244 & -27.8605 & -27.3154 & -28.4504 & -28.0865 & -28.0820 & -27.9490 & -27.7120 \\
2 & 1.0466 & 0.9980 & 3.0359 & 2.7390 & 3.4086 & 3.3403 & 3.2652 & 3.2313 & 2.9916 \\
3 & 60.0314 & 59.9334 & 57.6471 & 56.4789 & 57.8310 & 57.1236 & 57.7789 & 57.5223 & 58.6009 \\
4 & -51.4155 & -51.4784 & -48.1167 & -48.4238 & -48.5093 & -48.6576 & -47.6951 & -47.7375 & -46.9516 \\
5 & 17.9118 & 17.8795 & 18.8580 & 19.7368 & 17.7844 & 18.0040 & 17.7863 & 17.9568 & 17.7619 \\
6 & 5.0420 & 4.9948 & 5.5417 & 5.4296 & 5.4849 & 5.4645 & 5.6128 & 5.6664 & 5.8271 \\
7 & -21.1914 & -21.1773 & -23.6594 & -23.3149 & -24.0917 & -23.8679 & -23.7066 & -23.5907 & -23.3152 \\
8 & 20.9325 & 20.9482 & 19.0364 & 19.2070 & 18.0401 & 18.0877 & 18.0809 & 18.1434 & 17.8679 \\
9 & -6.9226 & -6.9081 & -8.2664 & -8.1840 & -8.7051 & -8.5336 & -8.3421 & -8.2612 & -8.2027 \\
10 & 11.2333 & 11.2367 & 10.6169 & 10.5831 & 10.5833 & 10.5517 & 10.6161 & 10.6306 & 10.6710 \\
11 & -5.1461 & -5.1598 & -6.3676 & -6.5982 & -6.2110 & -6.1491 & -6.0316 & -5.9587 & -5.8508 \\
12 & 6.3431 & 6.2513 & 7.3043 & 6.9409 & 7.7895 & 7.7302 & 7.8220 & 7.8406 & 8.2621 \\
13 & 92.9299 & 92.9608 & 90.3733 & 91.0558 & 87.3872 & 87.4771 & 86.9468 & 87.0484 & 86.0385 \\
14 & 13.3612 & 13.5036 & 16.2501 & 17.4525 & 15.0792 & 15.9671 & 14.7482 & 15.0328 & 13.6140 \\
15 & 3.0707 & 3.1065 & 1.8050 & 2.3636 & 0.8853 & 1.1887 & 1.0572 & 1.1947 & 1.2109 \\ \hline
\end{tabular}
\caption{Reaction energies computed with the a$X$Z basis sets using Peterson's and our $\beta_\text{opt}$ and Q5 extrapolated value}
\end{table}

\begin{table}[]
\begin{tabular}{r|rrrrrrr}
\hline
Rxn. & \multicolumn{2}{c}{DZ-F12} & \multicolumn{2}{c}{TZ-F12} & \multicolumn{2}{c}{QZ-F12} & CBS \\
 & $\beta_\text{ref}$ & $\beta_\text{opt}$ & $\beta_\text{ref}$ & $\beta_\text{opt}$ & $\beta_\text{ref}$ & $\beta_\text{opt}$ &  \\ \hline
1 & -24.7962 & -25.4801 & -27.5027 & -27.3256 & -27.7851 & -27.7570 & -27.7120 \\
2 & 3.4573 & 2.6088 & 3.4228 & 3.0806 & 3.2217 & 3.1492 & 2.9916 \\
3 & 59.6444 & 58.6599 & 57.3227 & 55.7603 & 57.4069 & 57.0134 & 58.6009 \\
4 & -48.0735 & -49.4117 & -47.5485 & -48.3344 & -47.5588 & -47.9977 & -46.9516 \\
5 & 14.8953 & 15.5299 & 17.6453 & 18.5343 & 18.2845 & 18.4280 & 17.7619 \\
6 & 7.0656 & 6.7892 & 5.6538 & 5.4537 & 5.6744 & 5.7019 & 5.8271 \\
7 & -20.3853 & -20.3840 & -23.1572 & -22.9782 & -23.4778 & -23.4180 & -23.3152 \\
8 & 19.4133 & 20.2251 & 18.7185 & 19.1717 & 18.3042 & 18.4085 & 17.8679 \\
9 & -5.8433 & -6.0238 & -7.7611 & -7.6200 & -8.2271 & -8.1608 & -8.2027 \\
10 & 10.4460 & 10.8235 & 10.6023 & 10.5892 & 10.5362 & 10.5695 & 10.6710 \\
11 & -4.6016 & -4.8258 & -5.7521 & -5.7389 & -5.9201 & -5.9028 & -5.8508 \\
12 & 8.3220 & 7.3500 & 7.4953 & 7.0031 & 7.8964 & 7.8111 & 8.2621 \\
13 & 85.9182 & 88.5086 & 87.7679 & 89.1863 & 87.5434 & 87.7553 & 86.0385 \\
14 & 11.1488 & 11.3700 & 15.5399 & 16.1721 & 15.0428 & 15.2624 & 13.6140 \\
15 & 1.8245 & 2.4727 & 1.1503 & 1.7258 & 1.3121 & 1.4154 & 1.2109 \\ \hline
\end{tabular}
\caption{Reaction energies computed with the $X$Z-F12 basis sets using Peterson's and our $\beta_\text{opt}$ and Q5 extrapolated values}
\end{table}

\begin{table}[]
\begin{tabular}{r|rrrrrr|rrrr}
\hline
 & \multicolumn{2}{r}{aDZ} & \multicolumn{2}{r}{aTZ} & \multicolumn{2}{r|}{aQZ} & \multicolumn{2}{r}{DZ-F12} & \multicolumn{2}{r}{TZ-F12} \\
 & $\beta_\text{ref}$ & $\beta_\text{opt}$ & $\beta_\text{ref}$ & $\beta_\text{opt}$ & $\beta_\text{ref}$ & $\beta_\text{opt}$ & $\beta_\text{ref}$ & $\beta_\text{opt}$ & $\beta_\text{ref}$ & $\beta_\text{opt}$ \\ \hline
\ce{H2O}$\dots$\ce{H2O} & -4.906 & -4.904 & -4.868 & -4.869 & -4.822 & -4.823 & -4.799 & -4.822 & -4.802 & -4.804 \\
\ce{HCCH}$\dots$\ce{HCCH} & -1.395 & -1.398 & -1.386 & -1.408 & -1.366 & -1.370 & -1.345 & -1.342 & -1.367 & -1.356 \\
\ce{CH3OH}$\dots$\ce{CH3OH} & -5.666 & -5.664 & -5.568 & -5.577 & -5.492 & -5.497 & -5.485 & -5.515 & -5.488 & -5.492 \\
\ce{H2O}$\dots$\ce{CH3OH} & -5.544 & -5.543 & -5.471 & -5.477 & -5.413 & -5.417 & -5.388 & -5.416 & -5.402 & -5.405 \\
\ce{HCCH}$\dots$\ce{H2O} & -2.943 & -2.943 & -2.843 & -2.854 & -2.803 & -2.803 & -2.760 & -2.771 & -2.779 & -2.771 \\ \hline
\end{tabular}
\caption{Interaction energies, kJ/mol}
\end{table}

\begin{table}[]
\begin{tabular}{rrrrrrr}
\hline
 &  & \ce{B2}$\rightarrow$ \ce{B2+} & \ce{C2}$\rightarrow$ \ce{C2+} & \ce{N2} $\rightarrow$ \ce{N2+} & \ce{O2} $\rightarrow$ \ce{O2+} & \ce{F2} $\rightarrow$ \ce{F2+} \\ \hline
DZ-F12 & $\beta_\text{ref}$ & 8.705 & 12.362 & 15.691 & 11.146 & 16.032 \\
DZ-F12 & $\beta_\text{opt}$ & 8.706 & 12.364 & 15.695 & 11.158 & 16.047 \\
TZ-F12 & $\beta_\text{ref}$ & 8.727 & 12.380 & 15.735 & 11.220 & 16.102 \\
TZ-F12 & $\beta_\text{opt}$ & 8.730 & 12.382 & 15.740 & 11.227 & 16.106 \\
QZ-F12 & $\beta_\text{ref}$ & 8.734 & 12.386 & 15.750 & 11.247 & 16.131 \\
QZ-F12 & $\beta_\text{opt}$ & 8.737 & 12.388 & 15.753 & 11.250 & 16.134 \\
5Z-F12 & $\beta_\text{ref}$ & 8.735 & 12.387 & 15.755 & 11.255 & 16.139 \\
5Z-F12 & $\beta_\text{opt}$ & 8.738 & 12.389 & 15.757 & 11.257 & 16.140 \\
aDZ & $\beta_\text{ref}$ & 8.695 & 12.352 & 15.638 & 11.044 & 15.940 \\
aDZ & $\beta_\text{opt}$ & 8.695 & 12.352 & 15.638 & 11.045 & 15.942 \\
aTZ & $\beta_\text{ref}$ & 8.731 & 12.383 & 15.732 & 11.212 & 16.095 \\
aTZ & $\beta_\text{opt}$ & 8.730 & 12.381 & 15.733 & 11.215 & 16.097 \\
aQZ & $\beta_\text{ref}$ & 8.736 & 12.388 & 15.751 & 11.247 & 16.130 \\
aQZ & $\beta_\text{opt}$ & 8.736 & 12.386 & 15.751 & 11.247 & 16.130 \\
a5Z & $\beta_\text{ref}$ & 8.736 & 12.388 & 15.756 & 11.255 & 16.139 \\
a5Z & $\beta_\text{opt}$ & 8.737 & 12.388 & 15.757 & 11.256 & 16.140 \\
CBS &  & 8.738 & 12.389 & 15.760 & 11.260 & 16.144 \\
\hline
\end{tabular}
\caption{Ionization Potentials, kJ/mol}
\end{table}

\begin{table}[]
\begin{tabular}{rrrrrrrrr}
basis &  & \ce{HF} & \ce{F2} & \ce{H2O} & \ce{N2} & \ce{NH3} & \ce{OH} & \ce{CH} \\ \hline
aDZ & $\beta_\text{ref}$ & -175.19 & -289.82 & -299.55 & -419.83 & -373.26 & -150.30 & -103.32 \\
aDZ & $\beta_\text{opt}$ & -175.27 & -290.00 & -299.69 & -420.19 & -373.44 & -150.38 & -103.35 \\
aTZ & $\beta_\text{ref}$ & -176.03 & -284.51 & -303.44 & -422.51 & -382.29 & -152.70 & -107.79 \\
aTZ & $\beta_\text{opt}$ & -176.10 & -285.04 & -303.68 & -423.31 & -382.74 & -152.90 & -107.98 \\
aQZ & $\beta_\text{ref}$ & -177.90 & -285.29 & -306.16 & -426.54 & -385.73 & -154.32 & -108.96 \\
aQZ & $\beta_\text{opt}$ & -177.87 & -285.25 & -306.20 & -426.70 & -385.86 & -154.36 & -109.00 \\
a5Z & $\beta_\text{ref}$ & -177.31 & -284.44 & -305.73 & -426.73 & -385.57 & -154.09 & -108.95 \\
a5Z & $\beta_\text{opt}$ & -177.40 & -284.54 & -305.89 & -427.06 & -385.82 & -154.19 & -109.03 \\
a6Z & $\beta_\text{ref}$ & -177.23 & -284.30 & -305.64 & -426.78 & -385.54 & -154.06 & -108.94 \\
a6Z & $\beta_\text{opt}$ & -177.29 & -284.37 & -305.77 & -427.08 & -385.76 & -154.13 & -109.02 \\
a7Z & $\beta_\text{ref}$ & -177.23 & -284.27 & -305.68 & -426.81 & -385.62 & -154.07 & -108.97 \\
a7Z & $\beta_\text{opt}$ & -177.27 & -284.31 & -305.77 & -427.06 & -385.77 & -154.13 & -109.03 \\
DZ-F12 & $\beta_\text{ref}$ & -172.19 & -278.97 & -296.43 & -415.48 & -372.76 & -148.78 & -104.21 \\
DZ-F12 & $\beta_\text{opt}$ & -173.31 & -281.81 & -298.18 & -418.67 & -374.47 & -149.67 & -104.49 \\
TZ-F12 & $\beta_\text{ref}$ & -175.56 & -283.16 & -302.46 & -422.87 & -381.32 & -152.26 & -107.68 \\
TZ-F12 & $\beta_\text{opt}$ & -176.05 & -284.18 & -303.44 & -424.40 & -382.56 & -152.79 & -108.13 \\
QZ-F12 & $\beta_\text{ref}$ & -176.73 & -283.96 & -304.80 & -425.50 & -384.27 & -153.58 & -108.52 \\
QZ-F12 & $\beta_\text{opt}$ & -177.02 & -284.31 & -305.37 & -426.42 & -385.15 & -153.88 & -108.82 \\
5Z-F12 & $\beta_\text{ref}$ & -176.99 & -284.09 & -305.29 & -426.34 & -385.03 & -153.87 & -108.73 \\
5Z-F12 & $\beta_\text{opt}$ & -177.21 & -284.29 & -305.66 & -426.92 & -385.62 & -154.07 & -108.95 \\ \hline
\end{tabular}
\caption{Atomization energies, kJ/mol}
\end{table}